\documentclass[9pt,conference]{IEEEtran}

\usepackage[preprint]{waspaa25}
\usepackage{bm} 
\usepackage{tikz}
\usepackage{pgfplots}
\usepackage{scalerel}
\usepackage{multirow}
\usepackage{makecell}
\usepackage{booktabs}
\usepackage{mathtools}
\usepackage{nicematrix}
\usepackage[caption=false]{subfig}
\usetikzlibrary{positioning}
\usetikzlibrary{tikzmark} 
\usetikzlibrary{arrows}   
\usetikzlibrary{calc}     
\tikzstyle{every picture}+=[remember picture]
\pgfplotsset{compat=1.7}

\usepackage{soul}

\title{Online incremental learning for audio classification \\using a pretrained audio model }

\name{Manjunath Mulimani, Annamaria Mesaros\thanks{This work was supported by Jane and Aatos Erkko Foundation under grant number 230048, "Continual learning of sounds with deep neural networks". \\
The authors wish to thank CSC-IT Centre of Science Ltd., Finland,  for providing computational resources.}}
\address{Signal Processing Research Centre, \textit{Tampere University}, Tampere, Finland \\
\{manjunath.mulimani, annamaria.mesaros\}@tuni.fi
}

\begin{document}
\maketitle

\begin{abstract}
Incremental learning aims to learn new tasks sequentially without forgetting the previously learned ones.
Most of the existing incremental learning methods for audio focus on training the model from scratch on the initial task, and the same model is used to learn upcoming incremental tasks. The model is trained for several iterations to adapt to each new task, using some specific approaches to reduce the forgetting of old tasks.
In this work, we propose a method for using generalizable audio embeddings produced by a pre-trained model to develop an online incremental learner that solves sequential audio classification tasks over time. Specifically, we inject a layer with a nonlinear activation function between the pre-trained model's audio embeddings and the classifier; this layer expands the dimensionality of the embeddings and effectively captures the distinct characteristics of
sound classes. 
Our method adapts the model in a single forward pass (online) through the training samples of any task, with minimal forgetting of old tasks. 
We demonstrate the performance of the proposed method in two incremental learning setups: one class-incremental learning using ESC-50 and one domain-incremental learning of different cities from the TAU Urban Acoustic Scenes 2019 dataset; for both cases, the proposed approach outperforms other methods.
\end{abstract}

\section{Introduction}
\label{sec:intro}

Incremental or continual learning is the subfield of machine learning in which 
a model must acquire new knowledge from continuously evolving audio data (often divided into distinct tasks/stages) over time and be performant on previously acquired knowledge. 
Specifically, an incremental learner or model learns from the current audio classification task in the absence of the data of previous tasks, leading to catastrophic forgetting \cite{parisi2019continual} of tasks learned previously. The storage of data from previous tasks may not be possible in all applications due to computational, copyright, or security constraints, making a complete retraining impossible when new knowledge is to be incorporated in a model.

Many incremental learning methods have been proposed and successfully applied to solve audio classification tasks in different experimental setups, e.g. task-incremental learning (TIL) \cite{muthuchamy2023adapter, karam2022task}, class-incremental learning (CIL) \cite{mulimani2024class, wang2022learning, wang2019continual},  
and domain-incremental learning (DIL) \cite{mulimani2024online, mulimani2024acoustic}.  
TIL is a task-aware setup that requires a task ID with an input sample to get results from the task-specific classifier during inference. CIL is a task-agnostic setup; it includes a single expanding classifier to accommodate new classes and predicts the label of the input sample without a task ID during inference. DIL deals with incremental tasks that have the same classes but come from different domains \cite{van2019three}.

In this work, our focus is on both CIL and DIL. In most existing works on CIL/DIL, the base model is trained from scratch for the initial task before learning incremental tasks. The base model is then trained on incremental tasks either offline for several iterations \cite{mulimani2025domain, mulimani2025closer} 
or online by going through training data only once \cite{mulimani2024online}, using some specific approaches to reduce forgetting. 
The existing offline/online DIL approaches for audio are based on domain/task-aware setups which require a domain/task ID to select domain-specific features for classification during inference. The performance of these methods is poor in a domain-agnostic setup \cite{mulimani2025domain}. 

Recently, many pre-trained audio models such as PANNs CNN14 \cite{kong2020panns}, AST \cite{gong2021ast}, PaSST \cite{koutini2022efficient}, and SSAST \cite{gong2022ssast} are available in the literature. The audio representations produced by these models have shown excellent performance on various downstream machine listening tasks such as acoustic scene classification \cite{hao2024enhancing}, multi-label audio classification \cite{dinkel2024ced}, and sound event detection \cite{shao2024fine}.  
A few works use pre-trained models  for incremental learning of audio classification in the TIL setup \cite{karam2022task, muthuchamy2023adapter}. In \cite{muthuchamy2023adapter}, task-specific adapters are used with the pre-trained model, and in \cite{karam2022task}, a reply buffer is used to store a small number of samples from previously 
seen tasks to reduce forgetting. However, in both of these cases, either all the parameters of the pre-trained models are tuned in a full fine-tuning approach, or parameters of the task-specific adapters are tuned for Parameter-Efficient Transfer Learning (PETL); this parameter tuning is done for several iterations to adapt to the incremental audio tasks. In a different approach, the few-shot class-incremental learning methods \cite{wang2021few, li2023few, li2024few} use a few samples to adapt to the new incremental tasks. 

This work aims to combine incremental learning with a powerful pre-trained audio model. Specifically, we consider a fixed pre-trained model as a strong generic feature extractor and use its feature representations (embeddings) to adapt to any number of incremental downstream audio classification tasks. One of the easiest methods is to train the linear probe classifier on top of the pre-trained feature embeddings to learn incremental tasks. A linear probe is widely used to compute the generalization ability and discriminability of the pre-trained models \cite{alain2016understanding, liang2022simple}. However, as explained earlier, training a linear probe incrementally on a new task in the absence of data from the previous tasks leads to catastrophic forgetting. 

In this work, we propose an online incremental learning approach for audio classification using a pre-trained model, in which 
we inject a layer with a nonlinear activation function between the pre-trained audio model's embeddings and the classifier; this layer expands the dimensionality of the embeddings. We then compute a class prototypes-based weight matrix for classification. This proposed approach adapts to a new task in a single forward pass through its training data, with minimal forgetting of the previous tasks. 
This method relies on a  
RanPAC \cite{mcdonnell2023ranpac}, a method proposed for incremental learning of images. In comparison to \cite{mcdonnell2023ranpac}, we do not fine-tune the pre-trained model on a base/first task; 
instead, we aim to investigate the effectiveness of feature representations from a fixed pre-trained audio model for the incremental learning of the audio classification task. Therefore, our method is free from base task training, and is fully online: all tasks require only a single forward pass through the training data.  We use a CNN-based pre-trained audio model that produces a high-dimensional feature vector, therefore, design choices will change as per the complexity of the audio data and the audio pre-trained model.  We conduct detailed experiments to show the impact of each design choice of the proposed method for the incremental learning of audio classification tasks.

The contributions of this work are as follows: (1) we propose a unified framework for both task-agnostic CIL and DIL for audio classification; (2) we show the effectiveness of feature representations from a fixed pre-trained audio model in incremental learning of audio classification tasks; (3) the proposed approach is free from an initial base task training and it is directly used to learn incremental tasks using embeddings from the pre-trained model; (4)  we perform a detailed analysis on the effect of expansion of the dimension of the audio embeddings with nonlinear activation function in both CIL and DIL setups.

The rest of the paper is organized as follows: Section 2 presents the incremental CIL and DIL tasks setups, notations and the proposed online incremental learning method for audio classification. Section 3 introduces the datasets, training setup, baselines, implementation details, evaluation metrics, and results of CIL and DIL. Finally, conclusions are given in Section 4.

\section{Method}
\label{sec:method}
\subsection{Incremental tasks setup and notations}
\label{sec:setup}
In our incremental learning setup, a sequence of $T$ supervised audio classification tasks $\mathcal{D}_1,\cdots,\mathcal{D}_T$ is introduced to the model for learning in incremental time steps. A task $\mathcal{D}_t$ is composed of audio samples and corresponding class labels. The data from previous tasks is inaccessible at any current stage. This work considers both class-incremental learning (CIL) and domain-incremental learning (DIL) protocols. In CIL, disjoint class labels are present in each task, while in DIL, the same set of classes is present in all the tasks, but there is a domain/distribution shift among the audio samples present in each task.

We denote the total number of training samples in each task as $M_t$ and denote the total number of classes in all $T$ tasks as $C$. $\mathbf{x}_{t, m}$ and $\mathbf{y}_{t, m}$ are the m-th training sample and corresponding one-hot encoded label of length $C$ in a task $\mathcal{D}_t$. $\mathbf{f}_{t, m} \in \mathbb{R}^H$ and $\mathbf{f}_{test} \in \mathbb{R}^H$ are embeddings extracted from a frozen pre-trained model $\mathcal{P}$ for input training sample $\mathbf{x}_{t, m}$ and test sample $\mathbf{x}_{test}$, respectively.

\begin{figure}[htbp!]
  \centering
  \subfloat[]{\includegraphics[width=0.85\linewidth]{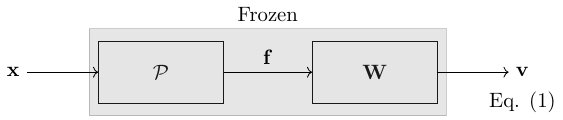}\label{Fig:stage1}}\\
  \subfloat[]{\includegraphics[width=0.85\linewidth]{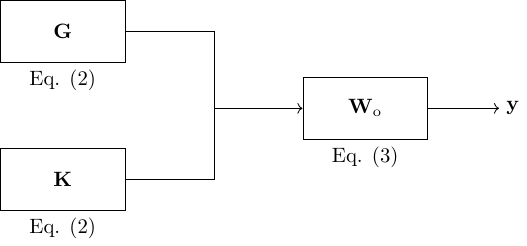}\label{Fig:stage2}}
 \caption{Overview of the proposed method. (a) Extracted features $\mathbf{f}$ from a frozen pre-trained model $\mathcal{P}$ for given input samples $\mathbf{x}$ are projected into a higher-dimensional space using frozen random weights $\mathbf{W}$, followed by a nonlinear activation function. (b) Gram ($\mathbf{G}$) and class prototypes ($\mathbf{K}$) matrices are iteratively updated and used to compute $\mathbf{W}_\text{o}$ by matrix inversion at each incremental task for prediction of classes seen so far.} 
        \label{Fig:Method}
\end{figure}

\subsection{Online incremental audio classification}
An overview of the proposed online incremental method for audio classification is shown in Fig.~\ref{Fig:Method}. 
The embedding features $\mathbf{f}_{t, m}$ of a given input training sample 
are projected into dimension $Q$ (typically, $H < Q$) using random projections $\mathbf{W} \in \mathbb{R}^{H\times Q}$, followed by an element-wise nonlinear activation function $\psi$ (e.g., ReLU) in each task  as:
\begin{equation}
    \mathbf{v}_{t, m} = \psi(\mathbf{f}_{t, m} ^\top\mathbf{W}), \quad \mathbf{v}_{test} = \psi(\mathbf{f}_{test} ^\top\mathbf{W})
    \label{eq1}
\end{equation}
where $\mathbf{v}_{t, m}$ and $\mathbf{v}_{test}$ are new features of length $Q$ in training and inference phases. $\mathbf{W}$ contains random weights rather than learned (training-free), generated once and kept frozen in all incremental learning stages. Further, we update the $Q \times Q$ Gram matrix $\mathbf{G}$ and $Q \times C$ class prototypes (CPs) matrix $\mathbf{K}$ iteratively in each task as:
\begin{equation}
    \mathbf{G} = \sum_{t=1}^T\sum_{m=1}^{M_t}\mathbf{v}_{t, m}\otimes\mathbf{v}_{t, m}, \quad \mathbf{K} = \sum_{t=1}^T\sum_{m=1}^{M_t}\mathbf{v}_{t, m}\otimes\mathbf{y}_{t, m}.
    \label{eq2}
\end{equation}
Usually, CPs are obtained by averaging the features of class $y$. However, in this work, $\mathbf{K}$ includes a CP for class $y$, denoted as $\mathbf{k}_y$ without averaging. In CIL, fewer classes than $C$ are available at the intermediate incremental stages, whereas in DIL, all $C$ classes are available to update $\mathbf{K}$. The $\mathbf{K}$ may generate highly correlated CPs, which results in a poor performance.
To generate more distinct CPs,  
we compute the decorrelated weight matrix using $\mathbf{G}$  and $\mathbf{K}$ as:
\begin{equation}
    \mathbf{W}_\text{o}=(\mathbf{G}+\lambda\mathbf{I})^{-1}\mathbf{K},
    \label{eq3}
\end{equation}
where $(\mathbf{G}+\lambda\mathbf{I})^{-1}$ is the $l_2$ regularized (or ridge regression based) inverse of the $\mathbf{G}$. $\mathbf{I}$ is the identity matrix and $\lambda$ is the ridge regression parameter. $\mathbf{W}_\text{o}$ can be considered as decorrelated CPs of size $Q \times C$. We update  $\mathbf{G}$, $\mathbf{K}$, and $\mathbf{W}_\text{o}$ matrices by seeing training samples only once, in the forward pass.
During inference, we use weights $\mathbf{W}_\text{o}$ to compute predictions of a class label of an input sample as $\mathbf{y}_{test}=\mathbf{v}_{test}\mathbf{W}_\text{o}$

We optimize the $\lambda$ for each task as follows. 
$\mathbf{G}$ and $\mathbf{K}$ are updated for each value of $\lambda$ in $\{10^{-8}, 10^{-7},\cdots, 10^{8}\}$ using randomly chosen 80\% of the training data $\mathbf{v}_{t}$ of task $t$. We then compute $\mathbf{W}_\text{o}$ using Eq. (\ref{eq3}) for prediction. We select the $\lambda$ that gives the minimum mean square error between targets and predictions from the remaining 20\% of the training data of the current task.

\begin{figure*}[!tbp]
  \centering
  \subfloat[]{\includegraphics[width=0.24\linewidth, height=0.22\linewidth]{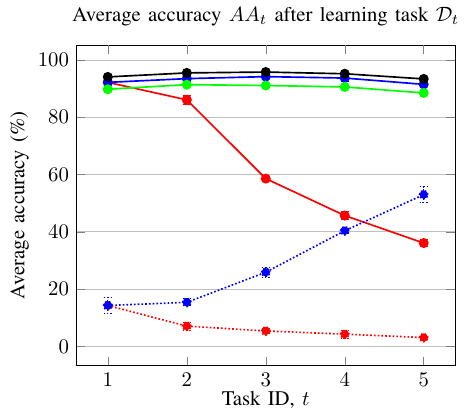}\label{fig:cil_accuracy}}
  \hfill
  \subfloat[]{\includegraphics[width=0.24\linewidth, height=0.22\linewidth]{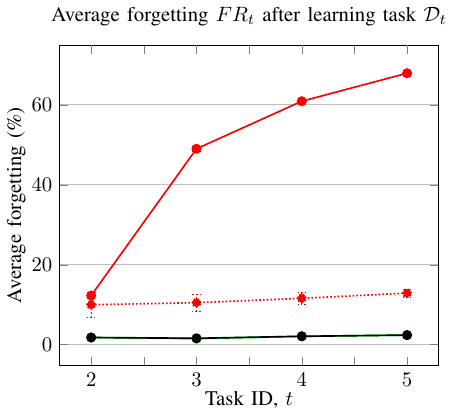}\label{fig:cil_forgetting}}
  \hfill
  \subfloat[]{\includegraphics[width=0.24\linewidth, height=0.22\linewidth]{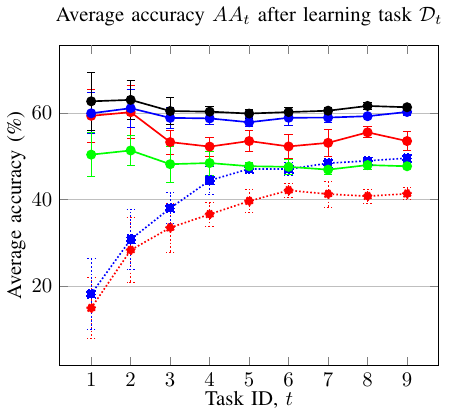}\label{fig:dil_accuracy}}
  \hfill
  \subfloat[]{\includegraphics[width=0.24\linewidth, height=0.22\linewidth]{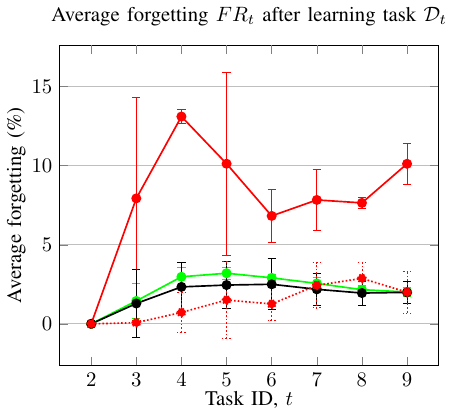}\label{fig:dil_forgetting}}\\
  \subfloat{\includegraphics[width=17cm]{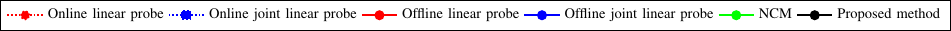}}
 \caption{Average accuracy and forgetting of the methods after learning the task $\mathcal{D}_t$. Average accuracy (a) and forgetting (b) of the current $\mathcal{D}_t$ and previously seen tasks in a CIL setup; Average accuracy (c) and forgetting (d) of the current $\mathcal{D}_t$ and previously seen tasks in a DIL setup.}
        \label{Fig:cil_dil}
\end{figure*}

\section{Evaluation and Results}
\label{sec:results}

\subsection{Datasets, training setup, baselines and evaluation metrics}
\label{datasets}

For \textbf{class-incremental learning,} we use the ESC-50 dataset \cite{piczak2015dataset}, which includes 50 environmental sound classes such as ``Cat", ``Footsteps", and so on. There are 40 5-second audio clips for each class, resulting in a total of 2000 clips divided into 5 folds. Each fold contains 8 samples per class; we select samples for training, validation, and testing with a ratio of 24 (3 folds):8 (1 fold):8 (1 fold) following \cite{piczak2015environmental}. The same splits are used in all experiments.
We distribute the 50 classes equally in 5 tasks, i.e., $T=5$, and each task includes 10 random disjoint classes. The model learns each task incrementally, and the performance of the model is evaluated on classes learned so far after learning the current task.

For \textbf{domain-incremental learning,} we use audio clips recorded in 9 different geographical locations: Barcelona, Helsinki, Lisbon, London, Lyon, Paris, Stockholm, Vienna, and Prague,  from the TAU Urban Acoustic Scenes 2019 development dataset \cite{Mesaros2018_DCASE}. Each location includes the same 10 acoustic scenes, such as ``Airport", ``Park", and so on. We use 20\% of the official training split as a validation set. 
We consider each location as a domain, or a task, resulting in 9 independent tasks, $T=9$. The model learns each task incrementally in random order, and the performance of the model is evaluated on domains learned so far after learning the current task.

We compare the performance of the proposed method with three different methods used to solve the same problem: (1) a linear probe classifier is trained on each incremental task using embeddings extracted from a pre-trained model for the current task data, (2) a joint linear probe classifier is trained using embeddings of all the tasks seen so far; this joint training violates the incremental learning setup, but it is given for completeness, (3) a Nearest Class Mean (NCM) classifier is obtained from constructing CPs by averaging the feature embeddings over training samples within classes. During inference, the class of a test sample is predicted using the highest cosine similarity between its feature embeddings and the set of CPs.

\subsection{Implementation details and evaluation metrics}
\label{metrics}

The 2048-dimensional feature vectors are extracted for data in each task using the PANNs CNN14 \cite{kong2020panns} pre-trained model. The log mel spectrograms are computed from 32 kHz resampled input audio recordings using default values given in \cite{kong2020panns}. The mini-batch size is set to 32, and we do not use any data augmentation. ReLU is used as a nonlinear activation function in Eq. (\ref{eq1}). The dimension $Q$ is set to 8192 for both CIL and DIL based on an ablation study with different values of $Q$, presented later in the paper.
For the baseline methods, linear and joint linear probes are trained using the cross-entropy loss, Adam optimizer \cite{loshchilov2017sgdr} with a learning rate of 0.0001 and the CosineAnnealingLR \cite{loshchilov2017sgdr} scheduler. The NCM and the proposed method do not require Adam-based weight updates.
The number of epochs is set to 1 and 100 to train linear and joint linear probes in online and offline settings, respectively. However, we stop the training in an offline setting using early-stopping criteria on the validation set.

Following the standard practice in incremental learning \cite{mulimani2025domain, mcdonnell2023ranpac}, we evaluate the performance of the method after learning the current task $\mathcal{D}_t$ using average accuracy and forgetting of all the tasks seen so far. Average accuracy is defined as:
\begin{equation}
    AA_t = \frac{1}{t}\sum_{j=1}^t ACC_{t, j},
\end{equation}
where $ACC_{t, j}$ is the accuracy of $j$-th task after learning the $t$-th task. Average forgetting is  defined as:
\begin{equation}
    FR_t = \frac{1}{t-1}\sum_{j=1}^{t-1} \max_{\hat{t}\in\{1, 2, \cdots, t-1\}} (ACC_{\hat{t}, j}-ACC_{t, j}),
\end{equation}
where $ACC_{\hat{t}, j}$ is the maximum accuracy of the previously seen $j$-th task after learning the $\hat{t}$-th task. 
In ESC-50, 5-fold cross-validation accuracy in each task is used to compute the $AA_t$/$FR_t$.

We run each experiment 5 times to learn the current task with 10 different classes in each run for CIL, and to learn the 9 cities in a different order in every run for DIL. Mean and standard deviation over average accuracy values across the runs are used to report the results.

\subsection{Results}
\label{results}

\begin{table}[]
  \centering
 \caption{Final average accuracy ($AA_T$) and forgetting ($FR_T$) of the methods over all the tasks $T=5$ on 50 classes of ESC50 for CIL and $T=9$ on 10 classes of 9 cities of TAU Urban Acoustic Scenes 2019 for DIL.}    
   \begin{tabular}{l|cc|cc}
  \toprule
 \multirow{2}{*}{Method} & \multicolumn{2}{c}{CIL} & \multicolumn{2}{c}{DIL} \\
 & $AA_T$ & $FR_T$ & $AA_T$ & $FR_T$\\
 \midrule
 \multicolumn{2}{l}{Online}\\
 \midrule
 LP & 03.1 $\pm$ 0.4  & 12.9 $\pm$ 1.1 & 41.4 $\pm$ 1.5  & 02.1 $\pm$ 1.4 \\
 JLP &53.0 $\pm$ 2.2  &  &49.6 $\pm$ 0.7 &\\
\midrule
\multicolumn{2}{l}{Offline}\\
 \midrule
 LP & 36.1 $\pm$ 1.3 & 67.9 $\pm$ 0.9 & 53.5 $\pm$ 2.3 & 10.2 $\pm$ 1.3 \\
 JLP & 91.5 $\pm$ 0.8 &&  60.3 $\pm$ 0.7 & \\
\midrule
NCM & {88.5 $\pm$ 0.0} & 02.4 $\pm$ 0.0 & 47.7 $\pm$ 0.0 & 02.1 $\pm$ 0.3 \\
 Proposed & \textbf{93.4 $\pm$ 0.1}  &\textbf{02.5 $\pm$ 0.1} & \textbf{61.4 $\pm$ 0.1}  &\textbf{02.0 $\pm$ 0.6}\\
 \bottomrule
\multicolumn{3}{l}{LP: linear probe, JLP: joint linear probe }
 \end{tabular}     
     \label{tab:cil_dil}
     \end{table}

We compare the performance of the proposed approach with baselines after learning all tasks for CIL ($T=5$) and DIL ($T=9$) in Table \ref{tab:cil_dil}. For a detailed analysis,  average accuracy $AA_t$ and forgetting $FR_t$ of the methods after learning the current task $\mathcal{D}_t$ are given in Fig.~\ref{Fig:cil_dil}.

\textbf{Class-incremental learning:} As expected, baseline methods trained offline perform better than online (1 epoch) training. The linear probe does not have access to the data of previous tasks. When it is trained only on the current task's data for several iterations, previously acquired knowledge on old classes is overwritten, resulting in increased average forgetting in Fig.~\ref{fig:cil_forgetting} and reduced average accuracy in Fig.~\ref{fig:cil_accuracy} as the model learns a new task.  Both the NCM classifier and the proposed approach are prototype-based methods and perform a single forward-only pass through the training samples of the current task. However, NCM fails to match the performance of the joint linear probe.
It may be because NCM suffers from high overlap between true class similarities (i.e, similarities between samples' feature embeddings and CPs of samples' class label) and inter-class similarities (i.e, similarities between samples' feature embeddings and CPs of classes other than samples' class label) in an incremental setup.

The proposed approach updates the Gram matrix $\mathbf{G}$ and CPs matrix $\mathbf{K}$, followed by computation of decorrelated $\mathbf{W}_\text{o}$ for classification. The decorrelation reduces the overlap between true-class and inter-class similarities. It expands the feature embeddings to a higher-dimensional space, followed by a nonlinear activation function, which helps the proposed approach to capture complex nonlinear relationships in feature embeddings and improve the linear separability between classes. Therefore, the proposed approach outperforms the joint linear probe with a high average accuracy $AA_t$ after learning each task, and 1.9\%p (percentage point) of improved final average accuracy $AA_T$ after learning all tasks. A good incremental learner adapts well to the new task, i.e., plasticity, with minimal forgetting of the already seen tasks, i.e., stability. From Fig.~\ref{fig:cil_accuracy}, Fig.~\ref{fig:cil_forgetting}, and the CIL column of the Table \ref{tab:cil_dil},  it can be seen that the proposed approach balances stability and plasticity with the minimal average forgetting and higher average accuracy as compared to all other methods, which do not have access to previous data.

\textbf{Domain-incremental learning:} Unlike CIL, which has new classes in each task, the DIL model learns from the embeddings of the same 10 classes from different domains, here identified as cities.  The average accuracy $AA_t$ of the online linear probe shown in Fig.~\ref{fig:dil_accuracy}  keeps improving after learning from each task using previously acquired domain knowledge.
We also observed that the models' performance on the previous domain is improved after learning the current domain in a few cases. For instance, the accuracy of `Helsinki' is improved after learning `Stockholm', which may be due to similar acoustic characteristics between domains.

An online linear probe adapts to the new task by going through the training data of the task only once and suffers less from forgetting.  Forgetting of the linear probe can be seen in Fig.~\ref{fig:dil_forgetting} and in the DIL column of the Table \ref{tab:cil_dil}.  As explained earlier, $AA_t$/$AA_T$ of the offline linear probe improved further, but also suffers from higher $FR_t$/$FR_T$ due to domain shift between cities. The NCM suffers more from high overlap between true class similarities and inter-class similarities in the DIL setup. Its average accuracy is worse as compared to both offline linear and joint linear probes. 

Similar to CIL, the proposed approach outperforms the offline joint linear probe by 1.1\%p and balances stability and plasticity with the minimal average forgetting and higher average accuracy as compared to all other methods, which do not have access to previous data. Therefore, the proposed online incremental learning approach can be considered as a unified framework that is suitable for both CIL and DIL in realistic scenarios.  

\begin{figure}[!tbp]
  \centering
  \subfloat[]{\includegraphics[width=0.5\linewidth]{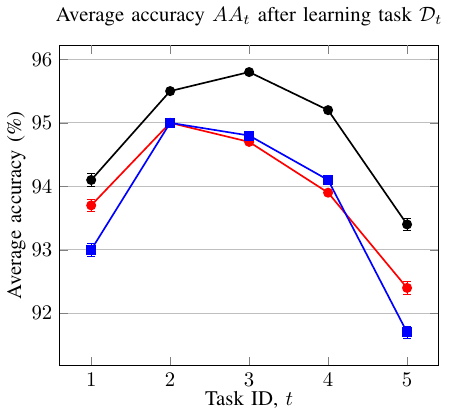}\label{fig:cil_ablation}}
  \hfill
  \subfloat[]{\includegraphics[width=0.5\linewidth]{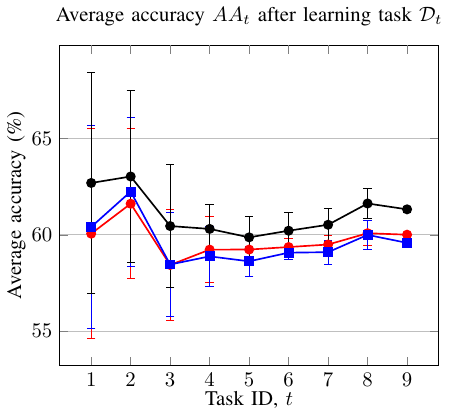}\label{fig:dil_ablation}}
  \\
  \subfloat{\includegraphics[width=7.0cm]{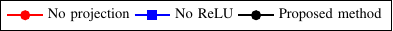}}
  \\
 \caption{Impact of the proposed method compared to alternatives without projection to $Q$ dimension using Eq. (\ref{eq1}) and without ReLU. (a) Average accuracy of CIL setup; (b) average accuracy of DIL setup.}
        \label{Fig:ablation}
     \vspace{-15pt}
\end{figure}

\begin{figure}[!tbp]
  \centering
  \subfloat[]{\includegraphics[width=0.5\linewidth]{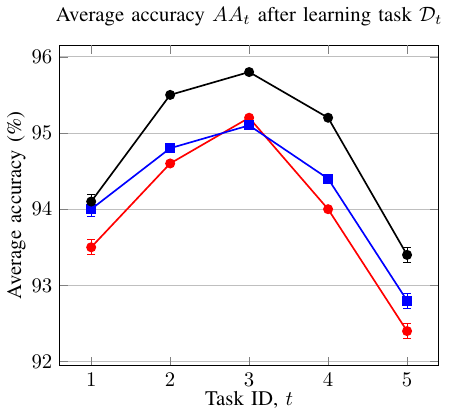}\label{fig:cil_q}}
  \hfill
  \subfloat[]{\includegraphics[width=0.5\linewidth]{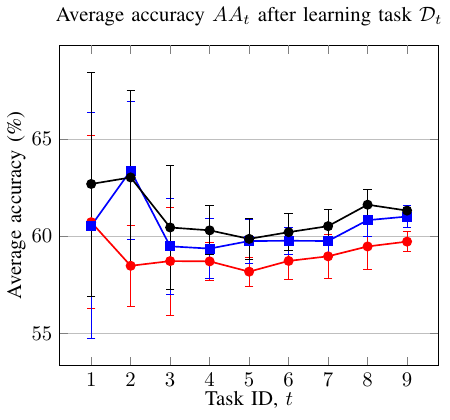}\label{fig:dil_q}}
  \\
  \subfloat{\includegraphics[width=6.0cm]{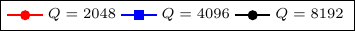}}
  \\
 \caption{Performance of the proposed method using different $Q$ dimensional values. (a) Average accuracy of CIL setup; (b) average accuracy of DIL setup.}
        \label{Fig:ablation_q}
\end{figure}

\textbf{Ablations:} Impact of projection of feature embeddings to $Q$ dimensions using frozen random weights $\mathbf{W}$, and use of ReLU nonlinear activation function followed by projection, are analyzed in Fig.~\ref{Fig:ablation} for both CIL and DIL setups. For the no projection experiment, we used the original 2048-dimensional embeddings to compute the Gram matrix $\mathbf{G}$, CPs matrix $\mathbf{K}$, followed by $\mathbf{W}_\text{o}$ for classification. Performance of the proposed approach without projection drops in both CIL (Fig.~\ref{fig:cil_ablation}) and DIL (Fig.~\ref{fig:dil_ablation}) setups.  However, projection without ReLU drops the performance even more. Therefore, nonlinearity followed by projection helps capture important information from the feature embeddings.

We analyze the impact of the dimension $Q$ on the performance of the proposed method in Fig.~\ref{Fig:ablation_q} for both CIL and DIL setups. Average accuracy improves as the dimension $Q$ increases, but
increasing $Q$ also leads to higher computational complexity, hence we choose to use dimension $Q=8192$ with the proposed method.
Expanding the dimension of the feature vector from $H=2048$ to $Q=8192$ increases the trainable parameters in the classifier by 4 times. For $C=50$ classes in the last phase of CIL, total trainable parameters in the joint linear probe are $H \times C$, i.e., 102400, and in the proposed method are $Q \times C$, i.e., 409600.  The $H \times Q$ parameters in the projection matrix $\mathbf{W}$ are frozen and not trainable. Furthermore, we need to update $Q \times Q$ Gram matrix G and $Q \times C$ CPs matrix only during training. However, these additional parameters are significantly less than the 80.8M parameters of the CNN14 model.

\section{Conclusion}
\label{sec:conclusion}

In this paper, we presented an online incremental learning approach for audio classification that starts from the PANNs CNN14 pre-trained model and builds upon its feature representations for incrementally learning audio tasks. 
Our proposed approach is suitable for both CIL and DIL and outperforms all other investigated methods in both cases. Because it relies on a pre-trained network and only needs to update a selected set of parameters in an online learning setting, the proposed method is suitable for realistic deployments. 
Future research includes investigating the performance of the proposed method with other audio pre-trained models.

\clearpage

\bibliographystyle{IEEEtran}
\bibliography{references}

\end{document}